\documentclass[9pt,twocolumn,twoside]{revtex4-1}%
\usepackage[top=3cm,bottom=3cm,right=2cm,left=2cm]{geometry}
\usepackage{graphicx}
\usepackage{verbatim}
\usepackage{amsmath}
\usepackage{amssymb}
\usepackage{mathrsfs}
\usepackage{hyperref}
\usepackage[title]{appendix}
\usepackage{mathtools} 

\newcommand{\bea}{\begin{eqnarray}}
\newcommand{\eea}{\end{eqnarray}}

\newcommand{\be}{\begin{equation}}
\newcommand{\ee}{\end{equation}}
\newcommand{\ba}{\begin{eqnarray}}
\newcommand{\ea}{\end{eqnarray}}

\newcommand{\NA}{N_{\rm A}}
\newcommand{\NC}{N_{\rm C}}
\newcommand{\NAi}{\tilde N_{\tilde{\rm A}}}

\begin{document}
\title{A robust transition to homochirality in complex chemical reaction networks}


\author{Gabin Laurent}
\email{gabin.laurent@espci.fr}
\author{David Lacoste}
\affiliation{Gulliver UMR CNRS 7083, ESPCI Paris, Universit\'e PSL, 75005 Paris, France}
\author{Pierre Gaspard}
\affiliation{Center for Nonlinear Phenomena and Complex Systems, Universit\'e Libre de Bruxelles, B-1050 Brussels, Belgium.}

\begin{abstract}
Homochirality, i.e. the dominance across all living matter of one enantiomer over the other among chiral molecules, is thought to be a key step in the emergence of life. Building on ideas put forward by Frank and many others, we proposed recently one such mechanism  in \href{https://doi.org/10.1073/pnas.2012741118}{G. Laurent et al., PNAS} (2021) based on the properties of large out of equilibrium chemical networks. We showed that in such networks, a phase transition towards an homochiral state is likely to occur as the number of chiral species in the system becomes large or as the amount of free energy injected into the system increases. This paper aims at clarifying some important points in that scenario, not covered by our previous work. We first analyze the various conventions used to measure chirality, introduce the notion of chiral symmetry of a network, and study its implications regarding the relative chiral signs adopted by different groups of molecules. We then propose a generalization of Frank's model for large chemical networks, which we characterize completely using methods of random matrices. This analysis can be extended to sparse networks, which shows that the emergence of homochirality is a robust transition.
\vspace{0.2cm}
\end{abstract}
\maketitle
\section{Introduction}
\label{sec_intro}

By definition a chiral molecule cannot be superposed to its mirror image, the molecule and its mirror image form a pair of enantiomers. A molecule can be chiral because it contains one or several asymmetric centers, or because its  tridimensional structure is itself chiral (for instance if the molecule has the shape of a helix).
The homochirality of life means precisely the dominance of only one enantiomer in this pair across the entire biosphere.
For instance, all amino acids are L-chiral and all sugars are D-chiral, and similarly all DNA and RNA molecules possess a specific handedness.
The origin for this selection of a single molecular handedness across all living matter is an outstanding problem in the research on the origins of life \cite{blackmond_origin_2010}. 
We do not know whether this property, which holds for the whole extent of life, must hold more generally and whether it is a prerequisite or a consequence of life \cite{brandenburg_homochirality_2020}. 
It would be important to know that, because, if it is a prerequisite of life, this could represent an efficient method to detect 
life on other planets than earth, and this would be also important for the design of artificial systems mimicking living systems. 

Modern enzymes rely on catalysis, which requires a tridimensional matching between the structure of the substrate or product with 
that of the active site on the enzyme. This tridimensional matching constrains the chirality of the enzyme with respect to that of its substrate or product. 
While this explains the relevance of chirality for biological catalysis, it remains difficult to imagine how homochirality could emerge in a prebiotic system, 
in which no enzyme was present initially. 

Many different mechanisms have been proposed over the years to explain homochirality: 
the parity violation in weak nuclear interaction for molecular energy levels \cite{quack_2002} or ionization by cosmic rays \cite{globus_chiral_2020}, or circularly polarized UV light from active star-forming region \cite{meierhenrich_amino_2008}. Another study suggests that the helical structure of biological molecules could be at the root of homochirality, because homochiral structures built of chiral amino acids are more thermodynamically stable than heterochiral structures \cite{liu_homochirality_2020}. 
While some of these mechanisms could possibly explain an initial chiral bias, these effects are often too small 
to explain the near complete homochirality of life, which needs to be amplified and maintained dynamically \cite{ribo_chemical_2019}.
	
The reaction network needs to be driven constantly out of equilibrium, otherwise the system would eventually converge
towards the racemic state, which is the thermodynamic equilibrium state \cite{PKBCA07}.	
The chemistry in this network also needs to be non-linear, so that complex dynamical behavior 
(instabilities, multistability, or oscillations) could exist. While not all these features need to be present, 
autocatalysis can bring the required non-linearities in the chemical network. 

Frank \cite{F53} and Kondepudi and Nelson \cite{KN83} have shown that 
certain non-equilibrium reaction networks based of this type can undergo a bifurcation towards a complete homochiral state. 
In Frank's model, each chiral monomer is amplified autocatalytically using achiral monomers as a resource  
and, in addition, there is a reaction of mutual decomposition between the two enantiomers, called chiral inhibition. 
In contrast to chiral inhibition, for which an experimental evidence was found \cite{joyce_chiral_1984}, there 
was no obvious way to implement experimentally the autocatalytic reactions considered by Frank. For this reason, it took chemists over forty 
years to find one experimental realization of Frank's model. Soai et al. discovered such a reaction in 1995 \cite{SH05}, a remarkable 
reaction but which is too specific to have played an important role in prebiotic chemistry. 
A possible reason for these difficulties to find suitable reaction schemes may be due to a lack of understanding of autocatalysis.
We now appreciate that autocatalysis needs not be direct, but could emerge as a 
collective property of a subpart of the network \cite{plasson_autocatalyses_2011} and we know how to identify it in the space of 
all possible chemical networks \cite{blokhuis_universal_2020}. As a result, autocatalysis appears to be 
much more common than previously thought.

In addition to this lack of suitable reaction schemes, there is also another more fundamental limitation in Frank's model, 
namely that it considers a small number of chemical species (three in the original model, one achiral species and 
two chiral species, enantiomers of each other). Only a few theoreticians realized this limitation, and explored generalization of this model involving many species. Among them, Sandars, who introduced an interesting polymerization model inspired by Frank's model \cite{sandars_chirality_2005} and Walker et al. who revisited that model a few years later \cite{gleiser_extended_2008}. Remarkably, these polymerization models contain a large number of species and are able to reach a full degree of homochirality only in the limit of very long polymers. 
	
In a recent work \cite{laurent_emergence_2021}, we argued for the need to consider large systems. There is indeed no reason to expect that the chemical composition was simple in the prebiotic world where homochirality first emerged. Primitive soups of any kind are more likely to have contained instead a myriad of strongly interacting organic species. The presence of a large number of chemical species or in the other words, the complex nature of prebiotic chemistry, is an essential feature, which ought to be included in models on the origins of life \cite{guttenberg_bulk_2017}. 

To support these ideas, we have quantified the abundance of chiral species 
in the space of all possible molecules using tools of chemoinformatics. We found that, as molecules grow in size, there is an explosion in the 
number of chemical species, both chiral and achiral. This is found in public databases containing millions 
of known molecules like PubChem \cite{pubchem_database} but also in artificial databases \cite{F07}, which are even bigger and contain in principle 
all possible molecules satisfying the basic rules of chemistry. Then, remarkably, one observes that, in both cases,  
the fraction of chiral compounds overtakes the achiral one when the length of molecules becomes of the order of 7-9 heavy atoms. 
This indicates that there is no need to consider very large molecules to observe a dominance of chiral species in the chemical space of all possible molecules: This should already happen with relatively simple and short molecules. 
A consequence of this cross-over is that the stereoisomer distribution goes from unimodal (with a maximum for achiral molecules) to bimodal (with two maxima corresponding to opposite enantiomers) as the size of molecules increases. This feature is a classic manifestation of a symmetry breaking mechanism.

Recently, a group of researchers led by Walker and Cronin found a similar crossover in a chemoinformatic study \cite{marshall_identifying_2021}. In that work, they introduced a measure for the complexity of a molecule, the molecular assembly index, which represents approximately the number of steps to make that molecule and which correlates with the molecular weight. In that study, they also confirmed that the high complexity molecules are predominantly chiral, as we observed in our work, which could possibly hint that the two cross-overs are related.

In our work  \cite{laurent_emergence_2021}, we  
showed that large non-equilibrium chemical networks are likely 
to undergo a chiral symmetry breaking precisely because they contain a large number of chiral species. 
This instability occurs irrespective of the nature and the abundance of chemical species, and 
irrespective of many details related to the topology of the network, 
except for the presence of a chiral symmetry. 
We analyze the chiral instability of molecular systems using a method based on random matrix theory, which has been first introduced to 
study of the stability of large complex systems, such as ecosystems or banking systems  \cite{allesina_stability_2012,may_robert_will_1972}.  
Since the mechanism behind this instability is very robust, we argue that homochirality could be  
a stereochemical imperative \cite{siegel_homochiral_1998}, which would be inevitable in sufficiently complex physico-chemical systems.

In this paper, we clarify a number of points related to our scenario on the emergence of homochirality in large molecular systems 
\cite{laurent_emergence_2021}. In particular,
we focus on the following questions:
Does the choice of a specific convention used to measure chirality matters for the emergence of homochirality? 
What features of the reaction network are required for the emergence of homochirality \cite{brandenburg_homochirality_2020}? 
In the next section, we recall the various conventions used to specify the chirality of chiral molecules. 
We then introduce a matrix framework to analyze the chirality of large chemical networks. We discuss a specific symmetry property of that matrix, which is related to an invariance under permutation in the designation D/L of the species. This symmetry has interesting consequences regarding the chiral signs of groups of molecules. We then introduce a random matrix technique to analyze the stability of chemical networks. We start with general considerations, and we then go towards more specific reaction schemes, such as generalizations of the Frank model for which we analyze the stability of the homochiral state. 
This part completes the conclusions of our previous paper, in which only the stability of the racemic state had been studied \cite{laurent_emergence_2021}.
We also discuss the role of chiral inhibition reactions and of the sparsity of the network before concluding on some main take home messages for this work. 

\subsection*{Chiral conventions}

In 1848, Pasteur discovered that salts of synthetic tartaric acid contained two distinct types of crystals, which are mirror images of each other. He went on to show that the two crystals rotate linearly polarized light in opposite directions and concluded that the racemic acid was made of two kinds of molecules with opposite optical activity.
Following this discovery, the measurement of optical activity became an important method to measure the chirality of a solution. When the rotation angle is clockwise, the substance is called dextrorotary and is denoted d(+), and if the angle is anticlockwise, the substance is called levorotary and is denoted l($-$). This (+) and ($-$) classification is important historically but it has many limitations in practice: (i) there is no way to determine the optical activity just by looking at the formula or the 3D structure of the molecule, (ii) the optical activity of chiral solutions also depends on the solvent, and (iii) the determination of the chirality from the optical activity becomes rather inaccurate as the number of chiral centers multiply. In fact, the only correlation which could be established between optical activity and chirality has been with molecules having a single chiral center, like amino acids or simple sugars. These simple sugars motivated the introduction of the Fischer-Rosanoff convention detailed below.

More commonly used in chemistry is the R/S descriptor of stereocenters (referring to Rectus and Sinister, Latin for right-handed and left-handed, respectively), where the ordering of the groups on each stereocenter is chosen based on the atomic numbers, according to a set of priority rules introduced by Cahn, Ingold, and Prelog in 1966 \cite{CIP_1966}. 
Once the substituents of a stereocenter have been assigned priorities according to these rules, the molecule is oriented in space so that the group with the lowest priority is pointed away from the observer. If the substituents are numbered from 1 (highest priority) to 4 (lowest priority), then the sense of rotation of a curve passing through 1, 2 and 3 distinguishes the stereocenters. A center with a clockwise sense of rotation is an R (rectus) center and a center with a counterclockwise sense of rotation is an S (sinister) center. 
The configuration R or S of a stereocenter can be easily determined by looking at the three-dimensional structure of the molecule. If the molecule has more than one stereocenter, its chirality is specified by the full list of R or S configurations for every one of its stereocenters. However, the R/S classification does not consistently maintain the ordering of the functional groups across, e.g., across all amino acids or all sugars, and there is no simple correlation between the R and S status of a molecular and its optical activity.

The D/L (named after Dexter and Laevus, Latin for right and left, respectively) convention is also known as the Fischer-Rosanoff convention.
Fischer introduced a planar representation of sugars with their carbon chain represented vertically and the most oxidated carbon atom on the top. 
Building on this notion, in 1906, Rosanoff chose glyceraldehyde, a monosaccharide, as the standard for denoting the stereochemistry of carbohydrates and other chiral molecules \cite{rosanoff_1906}.
Because Rosanoff did not know the absolute configuration of glyceraldehyde, he assigned it in a completely arbitrary manner:
He assigned the D (resp. L) prefix to (+)-glyceraldehyde (resp. ($-$)-glyceraldehyde), which in Fischer representation has the hydroxyl group (-OH)  attached to the chiral center on the right (resp. left) side of the molecule. Although Fischer rejected this nomenclature system, it was universally accepted and used to obtain the relative configurations of chiral molecules. The Fischer-Rosanoff convention allows to divide chiral molecules, such as amino acids and monosaccharides, into two classes, known as the D and L series. Note that the D/L system does not specify the sign of the rotation of plane-polarized light caused by the chiral molecule, nor does this convention correlates with R/S convention; it only correlates the configuration of the molecule in that convention with that of the glyceraldehyde. 

To summarize this section, all the conventions used in chemistry aimed to measure chirality are arbitrary, in fact the problem is not restricted to the identification of the chirality of molecules, there is no well defined way to assign handedness for any macroscopic chiral structure \cite{harris_molecular_1999}. We shall come back later to the consequence of the arbitrariness of chiral conventions.

\section{Symmetry and stability of large chemical networks}

\subsection{Kinetic equations and chiral symmetry}

\begin{figure*}[h]
\centering
\includegraphics[scale=0.40]{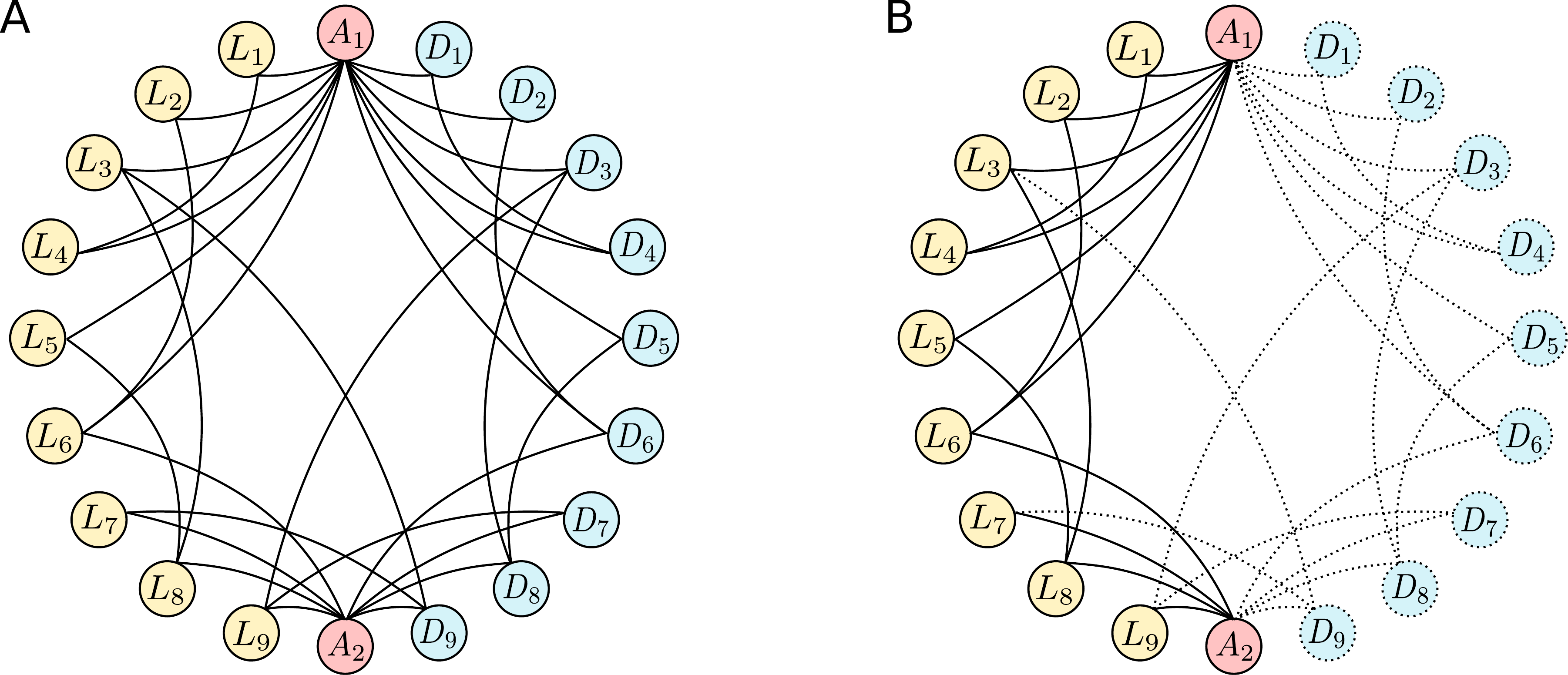}
\caption{
		Example of a reaction network containing two (resp. nine) achiral (resp. chiral) species. $A_i$ represents achiral species and $D_i$ (resp. $L_i$) represents the D (resp. L) enantiomer of the $i^{\rm th}$ chiral species. Solid lines connect random species involved in the same reaction. (A) Illustration of the chiral symmetry, according to which each chiral species linked by a random reaction is involved in the same - but mirrored - reaction involving mirrored species (enantiomers) with identical rate constants. As a result, the reaction network admits an axis of symmetry going through achiral species, here $A_1$ and $A_2$. (B) Illustration of the broken symmetry of a homochiral state. Dotted circles represent species that are no longer present in the system (or at extremely low concentrations) and dotted lines represent the reactions in which they were participating. Here, the simple case of a system undergoing a global symmetry breaking for all chiral species is depicted.}
\label{fig:CS_SB}
\end{figure*}

 We now investigate the properties of large non-equilibrium reaction networks, and in particular the conditions which are required to trigger a spontaneous mirror symmetry breaking towards a homochiral state \cite{HBAR17}.
Specifically, we consider a reaction network involving achiral and chiral species described by the concentration vector ${\bf c}$, which contains the vector ${\bf D}$ (resp. 
${\bf L}$) for the $N_{\rm C}$ D-enantiomers (resp. for the $N_{\rm C}$ L-enantiomers) and the vector ${\bf A}$ for the remaining $N_{\rm A}$ achiral species, as illustrated in Fig.~\ref{fig:CS_SB}.  This figure shows that the chiral symmetry 
is present in the initial racemic state of the reaction network (Fig.~\ref{fig:CS_SB}A), 
while the symmetry is broken in the homochiral state (Fig.~\ref{fig:CS_SB}B).  

As a paradigmatic example of an open chemical network, we consider a chemically stirred tank reactor (CSTR) \cite{blokhuis_reaction_2018}, which may be described by the following equations for the evolution of the concentrations: 
\be\label{kin_eqs}
\frac{d{\bf c}}{dt} = {\bf F}({\bf c}) + \frac{1}{\tau}({\bf c}_0-{\bf c}) \, , 
\ee
where ${\bf c}_0$ is the concentration vector of the species supplied from the environment at the rate $1/\tau$ and responsible for driving the system out of equilibrium and ${\bf F}({\bf c})={\boldsymbol\nu}\cdot{\bf w}({\bf c})$  are the reaction rates with specific chiral symmetry, 
which need not obey mass-action law. In this expression, ${\boldsymbol\nu}$ is the matrix of stoichiometric coefficients, ${\bf w}({\bf c})$ the set of net reaction rates (i.e., the difference between the forward and the backward reaction rates for a given reversible reaction). 
After the reaction has taken place, the species in excess flow out of the system at the same rate $1/\tau$ as for the supply, at the concentration ${\bf c}$, so that $\tau$ represents the mean residence time of the species in the system.

The trajectories of the dissipative dynamical system (\ref{kin_eqs}) are in general converging toward an attractor, which may be a steady state, a loop corresponding to periodic oscillations, or a strange attractor sustaining chaotic oscillations.  In the present context, the attractors may be assumed to be stationary and they may undergo bifurcations leading to multistability.

\onecolumngrid

\twocolumngrid

Symmetries play a central role in the equilibrium phase transitions of condensed matter physics, but in comparison they are more rare among biological systems. 
In view of the arbitrariness of the conventions on chirality mentioned above, we now consider the chiral symmetry of the kinetic equations (\ref{kin_eqs}).  The permutation of the concentrations of D and L enantiomers is carried out with
\be
{\boldsymbol{\mathsf S}}_{\bf c}\cdot{\bf c}=
\left( \begin{array}{ccc}
{\boldsymbol{\mathsf I}}  & 0 & 0\\
0 & 0 & {\boldsymbol{\mathsf I}}\\
0 & {\boldsymbol{\mathsf I}}  & 0 
\end{array}
\right) \cdot
\left( \begin{array}{c}
{\bf A} \\
{\bf D} \\
{\bf L}
\end{array}
\right) 
=
\left( \begin{array}{c}
{\bf A} \\
{\bf L} \\
{\bf D}
\end{array}
\right) ,
\ee
written in terms of the $N_{\rm S}\times N_{\rm S}$ matrix with $N_{\rm S}=N_{\rm A}+2N_{\rm C}$ and such that ${\boldsymbol{\mathsf S}}_{\bf c}^2 = {\boldsymbol{\mathsf I}}$, where ${\boldsymbol{\mathsf I}}$ denotes the corresponding identity matrix. Accordingly, the chiral symmetry of the kinetics can be expressed as
\be\label{chiral-sym-F}
{\bf F}({\bf c}) = {\boldsymbol{\mathsf S}}_{\bf c}^{-1}\cdot{\bf F}({\boldsymbol{\mathsf S}}_{\bf c}\cdot{\bf c}) \, .
\ee
We note that the concentration vector ${\bf c}_0$, which is controlling the supply of the reactor, may or may not satisfy the symmetry condition ${\boldsymbol{\mathsf S}}_{\bf c}\cdot{\bf c}_0={\bf c}_0$.  The breaking of chiral symmetry may happen spontaneously in the former case and explicitly by the external control in the latter case.  If the kinetic equations (\ref{kin_eqs}) admit a stationary solution ${\bf c}_s$, it will be either a racemic mixture if ${\boldsymbol{\mathsf S}}_{\bf c}\cdot{\bf c}_s={\bf c}_s$, or a steady state with some non-vanishing enantiomeric excess 
if ${\boldsymbol{\mathsf S}}_{\bf c}\cdot{\bf c}_s \ne {\bf c}_s$.

\subsection{Stability of steady states}

The stability of the racemic state can be studied  
by linearizing these equations about this state, which is defined by the condition ${\bf D}={\bf L}$. With the small parameter $\delta{\bf x}$, 
where ${\bf x}$ denotes the enantiomeric excess ${\bf x} \equiv \frac{1}{2} ({\bf  L}-{\bf D})$, we obtain
\be
\label{eq-M}
\frac{d}{dt} \delta{\bf x} =   {\boldsymbol{\mathsf M}}  \cdot\delta{\bf x} + \frac{1}{\tau}\, \delta{\bf x}_0 
\qquad\mbox{with}\qquad
{\boldsymbol{\mathsf M}}={\boldsymbol{\mathsf J}}-\frac{1}{\tau}\, {\boldsymbol{\mathsf I}}\, ,
\ee
where ${\boldsymbol{\mathsf J}}$ represents the Jacobian matrix
deduced from the kinetic equations~(\ref{kin_eqs}) and ${\boldsymbol{\mathsf I}}$ is the identity matrix. The racemic mixture is asymptotically unstable at long times if at least one of the eigenvalues of the matrix ${\boldsymbol{\mathsf M}}$ has a positive real part \cite{strogatz_nonlinear_1994}. We shall assume in the following that 
the chiral symmetry is not explicitly broken by the injection of species into the CSTR so that $\delta{\bf x}_0=0$.

If the steady state ${\bf c}_s$ is not a racemic mixture, the chiral symmetry implies the existence of another steady state of the kinetic equations (\ref{kin_eqs}), which is given by ${\boldsymbol{\mathsf S}}_{\bf c}\cdot{\bf c}_s$.  The linear stability of these steady states can be investigated by considering the linearized system
\be\label{lin-stab-gen}
\frac{d}{dt} \delta{\bf c} = {\boldsymbol{\mathsf N}}  \cdot\delta{\bf c} \, ,
\qquad\mbox{where}\qquad
{\boldsymbol{\mathsf N}}  =\frac{\partial{\bf F}}{\partial{\bf c}} - \frac{1}{\tau} \, {\boldsymbol{\mathsf I}}
\ee
is the $N_{\rm S}\times N_{\rm S}$ Jacobian matrix of the full system (\ref{kin_eqs}).  According to the chiral symmetry (\ref{chiral-sym-F}), both steady states have the same stability.

\subsection{Consequences regarding the relative signs adopted by different groups of molecules}

In the D/L system, naturally occurring amino acids are all L, while carbohydrates are nearly all D. In the R/S system, they are mostly S, but there are some common exceptions. Depending on the chosen convention, an enantiomer could be called X or its opposite. Therefore, the important point here is that all these conventions are arbitrary and the only possible meaning of homochirality is the dominance of one enantiomer over the other one, {\it independently of what happens for the other species}. This is the definition of homochirality we will assume in the following. Note that the average chirality of an ensemble of molecules is not particularly meaningful in this context, since it can not serve as an order parameter for the homochiral transition. 

In the literature on homochirality, many theoretical models have studied how a group of molecules can become collectively chiral (i.e., with the same sign) as a result of their interactions. Models of this kind have successfully explained the synchronization of groups of chiral molecules in a spatially extended system \cite{jafarpour_noise-induced_2017}. However, it does not follow from this mechanism that all chiral biomolecules should eventually adopt the same sign. Certain groups of chiral biomolecules appear to have a well defined sign while other groups have a different sign but understanding these relative signs is currently an open issue. 

In any case, our model differs significantly from previous models in that it incorporates 
from the start a crucial invariance with respect to permutation in the designation D/L of any member of a pair of enantiomers. Therefore, the model agrees with the possibility that different groups of chiral molecules adopt different signs of chirality as observed in today's life. There is no need for a global sign for all biological molecules, which anyway does not seem to exist.

Although we regard the sign of one given group as arbitrary since there is no unambiguous convention to fix it, we do not think that the relative signs adopted by different groups of molecules are random. In the case of the amino acids and the sugars for instance, these two groups of molecules probably work better together, when amino acids and sugars have opposite chirality. We suggest that the relative signs of different groups of molecules probably result from some kind of optimization linked with evolution. In this respect, these relative signs between groups of molecules could be similar to features of the genetic code, which are likely to have been optimized by evolution \cite{vetsigian_collective_2006}.

Naively, one could think that this difference in sign between amino acids and sugars is not really significant since one could always introduce another convention in which the two groups would have the same sign. With this kind of thinking, one would end up with a convention in which all biomolecules would have the same sign. However, such a convention can only be meaningful if the correlation is established between the chiralities of the two groups. The issue is thus to choose the most appropriate convention in an unambiguous way.
In the end, it seems to us that there might be some interest in redefining the chiral sign of molecules, in a way that correlates with the yield of metabolic reactions in which these molecules are participating. More precisely, one should put in the same chiral group, all molecules like sugars which work or are compatible with each other, and in a different group, molecules which work together but which are inhibited by or are incompatible with the first group \cite{ribo_spontaneous_2017}. In this way, one might find that all sugar molecules need to be in one group and all amino acids in a different one, as observed today in the D/L convention. By proceeding along the same lines for all the other groups of chiral biomolecules, one would end up with a coherent classification of biomolecules, which would be consistent with their metabolic or biological function. 

\subsection{Random matrix approach}

In this section, we present a random matrix approach to chemical reaction networks, which is built on  
ideas similar to those developed for ecosystems \cite{allesina_stability_2012,may_robert_will_1972,gardner_connectance_1970}. 
The application of random matrix theory to chemical reaction networks is not obvious, because  
 it would seem that such an application risks loosing all the specificities of the chemical
 reaction network. This turns out not to be so, as shown by the works done about ecosystems, because certain questions 
regarding the stability of networks are largely independent of the details contained in the chemical network.
The idea of using random matrices to analyze the stability of a complex system was pioneered by 
Robert May in his work on ecosystems \cite{may_robert_will_1972}. 
In that work, May assumed that there were no correlations in the elements of the matrix. More recently, the method 
has been extended by Tang et al. \cite{allesina_stabilitycomplexity_2015,allesina_stability_2012} to include correlations 
due to actual interactions among the species of the ecosystem. 
In addition to ecosystems, random matrix methods have also been used to analyze the stability
of large economical systems to which May also contributed \cite{moran_mays_2019}.

Let us now focus on chemical networks. In that case, the elements of the Jacobian matrix consists of products of rate constants, 
stoichiometric coefficients, and concentrations. 
The randomness of the elements of the Jacobian matrix 
come primarily from the randomness of rate constants and to a lesser extent from that of concentrations. 
In the specific case of enzymatic reactions, it has been shown that  
rate constants are distributed according to a log-normal distribution \cite{davidi_birds-eye_2018}. 
The general observation is that, in known reaction networks such as the Belousov-Zhabotinskii system, the rate constants may take very different values without regularity between them among the different reactions of the network.

Once one accepts the idea of treating the Jacobian matrix ${\boldsymbol{\mathsf J}}$  introduced in Eq.~(\ref{eq-M}) as a random matrix, 
	the simplest case is to also assume that the elements of this matrix are independent and identically distributed (i.i.d.) random variables
(but not necessarily Gaussian distributed) of mean value $\mu$ and variance $\sigma^2$ \cite{ginibre_statistical_1965}. 
Since the Jacobian matrix ${\boldsymbol{\mathsf J}}$ is in general depending on the concentrations of the racemic steady state, we note that both $\mu$ and $\sigma^2$ may vary with the mean residence time $\tau$, which is controlling the steady state, as shown in Supp.~Mat. of Ref. \cite{laurent_emergence_2021}.
When $\mu=0$, random matrix theory shows that the complex eigenvalues are uniformly distributed in a disk of radius $\sigma \sqrt{N_{\rm C}}$ 
in the limit of large values of $N_{\rm C}$ \cite{girko_circular_1985}. When $\mu \neq 0$, we find that there exists a single and isolated eigenvalue,
which is equal to $\mu N_{\rm C}$, and the corresponding eigenvector has uniform components to dominant order (cf. SI Appendix, section S2 in Ref. \cite{laurent_emergence_2021}). 
Two possible mechanisms for the instability of the racemic state then emerge for large $ N_{\rm C}$. 
Either (i) the instability occurs due to the isolated eigenvalue otherwise (ii) it occurs due to the eigenvalues located on the edge of the circle (which may be real or complex valued). It follows from this that when $\mu>0$ and $N_{\rm C}  \ge \max\{1/(\tau \mu), (\sigma/ \mu)^2 \}$, the system becomes unstable by the first mechanism where all species become simultaneously unstable, and when $(\sigma/\mu)^2 \ge N_{\rm C} \ge 1/(\tau \sigma)^2$, the system becomes unstable by the second mechanism and in this case only a subpart of all the species become unstable at the transition. In such cases, random matrix theory predicts that as $N_{\rm C}$ becomes large, these mechanisms of instability become more and more likely. 

There is however no reason to assume that elements of the Jacobian matrix of a reaction network should be in general i.i.d. random variables.
In fact, in our study \cite{laurent_emergence_2021}, we have observed correlations in the Jacobian matrix of a 
generalized Frank model, which arise from the difference between the diagonal and non-diagonal elements of the matrix.
In that model, eigenvalues do not fill a Girko circle but fill a domain of different shape, closer to an ellipse. Therefore, in our study of the 
generalized Frank model, we never observed mechanism (ii) but only mechanism (i).
This example shows that the elements of Jacobian matrices are in general correlated, but we proved that nevertheless 
certain features remain. In particular, if the matrix elements are statistically correlated, the non-dominant eigenvalues may have a different distribution, but the isolated eigenvalue behaves similarly. This means that the scenario (i) is expected to be robust.

Another interesting aspect of the random matrix approach is that it can describe important features of the network such as its sparsity.
The sparsity of the network could originate from the sparsity of the stoichiometric matrix ${\boldsymbol\nu}$, which has already been observed in studies of metabolism of living system, and which results from the low connectivity of certain species to the rest of the network. 
A second origin to sparsity lies in the fact that in a typical chemical system, the various rate constants can span many orders of magnitude, resulting in certain matrix elements becoming negligible compared to other ones. A third origin to sparsity may be due to the variability of certain concentrations species in the network, which could be large specially if the system is not well mixed.
Whatever its origin, the sparsity of the Jacobian matrix ${\boldsymbol{\mathsf J}}$ can be accounted for by the theoretical treatment of the previous section because it only affects the radius of the disk in which the eigenvalues are distributed in. More precisely, this radius which is $\sigma \sqrt{N_{\rm C}}$ for a non-sparse network becomes $\sigma \sqrt{\alpha N_{\rm C}}$ in a sparse matrix, where $\alpha \in [0,1]$ measures the percentage of non-zero elements in ${\boldsymbol{\mathsf J}}$. Moreover, in the case where $\mu \neq 0$, the isolated eigenvalue is approximately changed to $\mu N_{\rm C} \to  \alpha \mu N_{\rm C}$. These two changes implies that the criterion for the system to be unstable by the first mechanism becomes $N_{\rm C} \geq \max\{1/(\alpha \tau \mu), (\sigma / \mu \sqrt{\alpha})^2\}$, and by the second mechanism, $(\sigma / \mu \sqrt{\alpha})^2 \geq N_{\rm C} \geq 1/(\sigma \tau\sqrt{\alpha})^2$. In other words, the sparser the matrix ${\boldsymbol{\mathsf J}}$, the higher the number of chiral species needed for the chiral symmetry breaking to occur in a system with given $\mu, \sigma$, and $\tau$. In the end, if the number of chiral species is sufficiently high, the results remain unchanged.
An interesting consequence of that observation is that the sparsity coefficient of ${\boldsymbol{\mathsf J}}$ is a tuning parameter to control the homochiral transition at a fixed number of chiral species $N_{\rm C}$ (provided that $N_{\rm C}$ is in a range that allow the system to be unstable).

\section{Specific reaction schemes: Generalized Frank model}

In this section, we recall the construction of the generalized Frank model introduced in our previous work  \cite{laurent_emergence_2021},
and we then discuss an important point not studied in that work, 
which concerns the stability of the homochiral state. In order to test the random matrix scenario, 
we have introduced a generalization of Frank's model \cite{F53}, in which the numbers of chiral and achiral species have been significantly increased and we have assumed an arbitrary assignation L or D to each enantiomer. We also include reverse reactions in order to guarantee the compatibility with the existence of an equilibrium state even though the system is driven out of equilibrium. It is essential that the system be driven out of equilibrium in order for chirality to be maintained. We thus assume that the system is thermodynamically open, due to fluxes of matter in and out of the system.  

Let us also suppose that species entering the autocatalytic system are achiral but of high free energy, while the achiral species produced by the reactions involving the two D- and L-enantiomers have a lower free energy. 
In this regard,  the achiral species $\{A_a\}_{a=1}^{\NA}$ are of high free energy, and the achiral species $\{\tilde A_a\}_{a=1}^{\NAi}$ of low free energy.  The reaction networks are given by the following reactions:
\bea
&&{\rm A}_a +{\rm E}_i  \rightleftharpoons {\rm E}_j + {\rm E}_k \, , \label{react1} \\
&&{\rm A}_a +\bar{\rm E}_i  \rightleftharpoons \bar{\rm E}_j + \bar{\rm E}_k \, , \label{react2} \\
&&{\rm E}_i + \bar{\rm E}_j   \rightleftharpoons \tilde{\rm A}_b + \tilde{\rm A}_c  \, , \label{react3} 
\eea
where the enantiomer species are either ${\rm E}_m={\rm D}_m$ and $\bar{\rm E}_m={\rm L}_m$, or ${\rm E}_m={\rm L}_m$ and $\bar{\rm E}_m={\rm D}_m$ for 
each enantiomeric pair $m=i,j,k=1,2,\dots,\NC$. The achiral species are labeled with $a=1,2,\dots,\NA$; and $b,c=1,2,\dots,\NAi$.  Equations~(\ref{react1})-(\ref{react3}) define a total of $2^{N_{\rm C}-1}$ inequivalent reaction networks differing by the permutations of D- and L-enantiomers for 
some enantiomeric pairs.  For given reaction rates, all these networks manifest similar dynamical behaviors.  Among them, the network with ${\rm E}_m={\rm D}_m$ and $\bar{\rm E}_m={\rm L}_m$ for all the pairs $m=1,2,\dots,N_{\rm C}$ is the direct generalization of Frank's model, considered below. We note that the models (\ref{react1})-(\ref{react3}) all have some degree of enantioselectivity as in the original Frank model \cite{F53}.

As explained in Appendix C, the reversible generalized Frank model (\ref{react1})-(\ref{react3}) undergoes relaxation to racemic equilibrium in a closed system, corresponding to an infinite mean residence time $\tau=\infty$ in Eq.~(\ref{kin_eqs}).  Here below, we consider the generalized Frank model under far-from-equilibrium conditions where a bifurcation towards homochirality happens beyond a threshold in the nonequilibrium driving by the high free-energy achiral species $\{A_a\}_{a=1}^{\NA}$ with respect to the low free-energy ones $\{\tilde A_a\}_{a=1}^{\NAi}$.  In such far-from-equilibrium regimes, the reversed reactions are often playing a negligible role, which can be described by assuming that their reaction rate constants are arbitrarily small.  We show in Appendix C that this assumption is compatible with the existence of an equilibrium state.

\subsection{Condition for the existence of the fully homochiral state}

The argument developed in the previous section on random matrix theory is
general, and for that reason, one could get the impression that 
it applies to any fixed point, including the homochiral fixed point. 
In fact, it is not so because the chiral symmetry takes a specific form when applied 
in the racemic state. 

In the particular case when all the rate constants are equal, all the fixed points can be computed exactly and their stability can be characterized. We have found that there are indeed always some fixed points, which are attractors of the dynamics \cite{laurent_emergence_2021}.  
When rate constants are assumed to be random, and the Jacobian matrix is evaluated at the racemic state, where $\mu \ge 0$, one finds this state to be generically unstable by the mechanism (i), but that mechanism no longer works for homochiral states where one can have $\mu \le 0$. Therefore, homochiral fixed points can be stable. 
Our numerical study of the generalized Frank model confirmed this, because we never found that the instability of the racemic state could occur for sufficiently large $N_{\rm C}$ by the mechanism (ii). We only observed numerically that the instability occurred by the mechanism (i) in all conditions we explored.   
Only the mechanism (i) is robust with respect to the presence of correlations, but it is only relevant if $\mu >0$, while mechanism (ii) does not have this limitation 
but it is dependent on the assumptions of statistical independence of the elements.  
As mentioned above, Jacobian matrices are generically random but correlated, therefore, in all these real situations, only the first mechanism applies in full generality.

In this section, we analyze the stability of the homochiral state of the generalized Frank model more precisely. 
We define a D homochiral state such that the concentrations of all species $i$ satisfy $D_i \neq 0$ and $L_i = 0$.
This definition of homochiral state can be extended to the more general case where some species $L_i$ have a non zero concentration (with the corresponding concentration of $D_i$ at zero), thanks to the symmetry of the Jacobian matrix mentioned above.  

We recall that the generalized Frank model is described by the following $2+2\NC$ equations in the fully irreversible regime with $i,j,k=1..\NC$ \cite{laurent_emergence_2021}, when only one type of achiral species of high and low free energy are present (i.e. $\NA = \NAi = 1)$:

\begin{widetext}
\begin{eqnarray}
	\dot{A} &=& - \sum_{ijk\atop j\le k} k_{+ijk} \, A \, D_i  - \sum_{ijk\atop j\le k} k_{+ijk} \, A \, L_i + \frac{1}{\tau} (A_{0}-A) \, , \label{eq-A-num2} \\
	\dot{D}_m &=& -\sum_{ij\atop i\le j} k_{+mij} \, A \, D_m + \sum_{ij\atop m\le j} k_{+imj} \, A \, D_i  + \sum_{ij\atop j\le m} k_{+ijm} \, A \, D_i   -\sum_{i} \tilde k_{-mi} \, D_m \, L_i -\frac{1}{\tau} D_m \, , \label{eq-Dm-num2} \nonumber \\
	\dot{L}_m &=&-\sum_{ij\atop i\le j} k_{+mij} \, A \, L_m + \sum_{ij\atop m\le j} k_{+imj} \, A \, L_i + \sum_{ij\atop j\le m} k_{+ijm} \, A \, L_i -\sum_{i} \tilde k_{-im} \, D_i \, L_m -  \frac{1}{\tau} L_m \, , \label{eq-Lm-num2} \nonumber \\
	\dot{\tilde A} &=& 2\, \sum_{ij}  \tilde k_{-ij} \, D_i \, L_j - \frac{1}{\tau} \tilde A \, . \label{eq-Ai-num2} \nonumber
\end{eqnarray}
\end{widetext}

Note that equations (\ref{eq-A-num2}) are compatible with homochiral steady states because $\dot L_m=0$ for all $m$ is satisfied if $L_i=0$ for all $i$.
Using the condition that for all chiral species $i$ : $D_i \neq 0$ and $L_i = 0$ into the first equation of~(\ref{eq-A-num2}), we find an expression 
for the concentration of the activated achiral species $A$ in the steady state :
\begin{equation}
	A = \frac{A_0}{\tau \sum_{ijk \atop j \leq k} k_{+ijk} D_i +1} \, .
\end{equation}
The sum in the denominator can be simplified using the  central limit theorem (CLT) :
\begin{equation}
	\sum_{ijk \atop j \leq k} k_{+ijk} D_i  = \sum_i D_i \sum_{jk \atop j\leq k} k_{+ijk} 
	 =  \sum_i D_i \langle k_+ \rangle \frac{N_{\rm C} ( N_{\rm C} +1)}{2} \, .
\end{equation}
Thus, we find that $A$ reads 
\begin{equation}
	A = \frac{A_0}{\tau \langle k_+ \rangle \frac{N_{\rm C} ( N_{\rm C} +1)}{2} \sum_i D_i  +1} \, .
	\label{expression1}
\end{equation}
Summing the second equation in~(\ref{eq-A-num2}) over $m$, in the steady state, we find
\begin{equation}
\begin{multlined}
	-A\sum_m D_m \sum_{ij\atop i\le j} k_{+mij} + A\sum_i D_i \sum_{mj\atop m\le j} k_{+imj} \\
	+ A\sum_i D_i \sum_{mj\atop j\le m} k_{+ijm}-\frac{1}{\tau} \sum_m D_m = 0 \, .
\end{multlined}
\end{equation}
Using the CLT to perform the sums and the property that all $D_i$ are non zero and positive concentrations, we deduce  the alternative expression $A=A_0^*$ with
\begin{equation}
	A_0^* := \frac{2}{\tau \langle k_+ \rangle N_{\rm C} (N_{\rm C} + 1)} \, .
\label{A0threshold}
\end{equation}
Combining the two expressions for $A$, namely~(\ref{expression1}) and (\ref{A0threshold}), 
 we obtain an expression for $\sum_i D_i$ in the steady state
\begin{equation}
	\sum_i D_i = A_0 - \frac{2}{\tau \langle k_+ \rangle N_{\rm C} (N_{\rm C} + 1)} \, .
	\label{sum_D_i}
\end{equation}
Of course, a solution where $\sum_i D_i <0$ has no physical meaning here, therefore a full homochiral state exists only if $A_0 > A_0^*$ in terms of Eq. (\ref{A0threshold}).
This makes sense because this is precisely the condition for a spontaneous symmetry breaking to occur which we derived in Eq.~(6) of Ref.~\cite{laurent_emergence_2021}.
Note that in this derivation, we have used only the global condition $\sum_i D_i > 0$, 
which does not ensure that $D_i >0$, $\forall \, i$. In fact, going back to equation~(\ref{eq-A-num2}) and using once again the CLT, one finds that in the steady state,
\begin{eqnarray}
D_m &=& \frac{A \sum_ i D_i \big(\sum_{j\atop m\le j} k_{+imj}  + \sum_{j\atop j\le m} k_{+ijm} \big) }{A \sum_{ij\atop i\le j} k_{+mij} + \frac{1}{\tau}} \nonumber \\ &=& 
\frac{A \langle k_+ \rangle (N_{\rm C} +1) \sum_ i D_i }{A N_{\rm C} (N_{\rm C} +1) \langle k_+ \rangle/2 + 1/\tau} > 0 \, ,
\end{eqnarray}
which shows that all $D_i$'s are positive and approximately equal to each other since $\sum_i D_i > 0$ if $A_0 > A_0^*$. 

So far in this section, we have characterized the homochiral steady state. In the following section, we will study its stability. Finally, let us introduce a mean-field approximation that will be useful for the next part, such that $\forall i, D_i =D$. Then we obtain from Eq.~(\ref{sum_D_i}),
\be
D = \frac{A_0}{N_{\rm C}} - \frac{2}{\tau \langle k_+ \rangle N_{\rm C}^2 (N_{\rm C} +1)} = \frac{A_0 - A_0^*}{N_{\rm C}} \, .
\label{MF_D}
\ee
Note that a similar development can be done when multiple achiral species are present $N_{\rm A} > 1$. In this case, an arbitrary $N_A$ can be absorbed in the parameter $A_0$ in such a way that the threshold $A_0^*$ is trivially modified. An example of this rescaling is given in the Supp. Mat of \cite{laurent_emergence_2021}.
\subsection{Linear stability analysis of the fully homochiral state}

When using linear stability analysis about the homochiral state, the dynamics cannot be written in the form used for the racemic state  and Eq.~(\ref{lin-stab-gen}) should instead be considered
with the $(2N_{\rm C} + 1) \times (2N_{\rm C}+1)$ Jacobian matrix  given by
\be
	 {\boldsymbol{\mathsf N}} = \begin{pmatrix}
	\frac{\partial \dot{A}}{\partial A} & & \frac{\partial \dot{A}}{\partial {\boldsymbol{\mathsf D}}} & & \frac{\partial \dot{A}}{\partial {\boldsymbol{\mathsf L}}} \\
	\\
		\frac{\partial \dot{{\boldsymbol{\mathsf D}}}}{\partial A} & &\frac{\partial \dot{{\boldsymbol{\mathsf D}}}}{\partial {\boldsymbol{\mathsf D}}} & &\frac{\partial \dot{{\boldsymbol{\mathsf D}}}}{\partial {\boldsymbol{\mathsf L}}} \\
		\\
		\frac{\partial \dot{{\boldsymbol{\mathsf L}}}}{\partial A} & &\frac{\partial \dot{{\boldsymbol{\mathsf L}}}}{\partial {\boldsymbol{\mathsf D}}} & &\frac{\partial \dot{{\boldsymbol{\mathsf L}}}}{\partial {\boldsymbol{\mathsf L}}}
	\end{pmatrix} .
\ee
The elements of this matrix are detailed in Appendix~A  [Eqs.~(\ref{dapda})-(\ref{dlpdl})]. Since $\partial \dot{{\boldsymbol{\mathsf L}}}/\partial A$ and $\partial \dot{{\boldsymbol{\mathsf L}}}/\partial {\boldsymbol{\mathsf D}}$ are respectively a null vector and a null matrix in this homochiral state, the characteristic determinant of ${\boldsymbol{\mathsf N}}$ can be decomposed in the following determinant product of two diagonal square blocks:
\begin{equation}
\det\begin{pmatrix}{\boldsymbol{\mathsf N}}-\lambda{\boldsymbol{\mathsf I}}\end{pmatrix}= \det\begin{pmatrix}
	\frac{\partial \dot{A}}{\partial A}-\lambda& & \frac{\partial \dot{A}}{\partial {\boldsymbol{\mathsf D}}} \\ \\ 
	\frac{\partial \dot{{\boldsymbol{\mathsf D}}}}{\partial A} & &\frac{\partial \dot{{\boldsymbol{\mathsf D}}}}{\partial {\boldsymbol{\mathsf D}}} -\lambda{\boldsymbol{\mathsf I}}
\end{pmatrix}
\det\begin{pmatrix}
\frac{\partial \dot{{\boldsymbol{\mathsf L}}}}{\partial {\boldsymbol{\mathsf L}}} -\lambda{\boldsymbol{\mathsf I}} \, 
\end{pmatrix} .
\end{equation}
The spectrum of ${\boldsymbol{\mathsf N}}$ is thus determined by the eigenvalues of its upper-left block and its lower-right block. The spectrum of the block $\partial \dot{{\boldsymbol{\mathsf L}}}/\partial {\boldsymbol{\mathsf L}}$ and its maximal eigenvalue are given in appendix A. The spectrum of the upper-left block of matrix ${\boldsymbol{\mathsf N}}$ is determined in Appendix B. 

In the particular case $A_0 = 2A_0^*$, the eigenvalues of the latter submatrix become degenerate and equal to  $\lambda_+=\lambda_- = -1/ \tau$. Since it is always negative, the stability of the homochiral state is entirely determined by the maximal eigenvalue of $\partial \dot{{\boldsymbol{\mathsf L}}} / \partial {\boldsymbol{\mathsf L}}$. 

In the general case when $A_0 \neq 2A_0^*$, the eigenvalues are non-degenerate and read 
\be
\lambda_+ = -\frac{1}{\tau} \, ,  \qquad
\lambda_- = \frac{1}{\tau}\left(1-\frac{A_0}{A_0^*}\right) \, . \label{lambda_m}
\ee
The first eigenvalue $\lambda_+$ is always negative. The second one $\lambda_-$ is negative when $A_0 > A_0^*$, i.e., when the racemic state becomes unstable and at the same time, a  physically acceptable homochiral state with $D_i >0$, $\forall i$ becomes possible. As shown in Appendix~A, the maximal eigenvalue of the matrix $\partial \dot{{\boldsymbol{\mathsf L}}} / \partial {\boldsymbol{\mathsf L}}$ is
\be
\lambda_{\rm max}^L = - N_{\rm C} D \langle \tilde{k}_- \rangle
\label{lambda_L-main}
\ee
in the large $N_{\rm C}$ limit, where $\langle \tilde{k}_- \rangle$ is the mean value of chiral inhibition rate constants. Furthermore, in the mean-field approximation, we have $D~=~(A_0 - A_0^*)/N_{\rm C}$ by Eq.~(\ref{MF_D}). Thus, in the regime of the homochiral state (i.e., when $A_0 > A_0^*$), this state is stable because $\lambda^L_{\max} < 0$. 

To summarize, when $A_0 > A_0^*$ the racemic state is unstable but the homochiral state is an attractor of the dynamics. Conversely, when $A_0 < A_0^*$, the unphysical condition $D < 0$ which is obtained means that there can be no stable homochiral state. In that case, the system should converge towards the racemic state because it is an attractor of the dynamics.
The presence of the mean value of chiral inhibition rate constants in $\lambda_{\rm max}^L $ underlines
its key role in the stability of the homochiral state. 
We have tried a number of reaction schemes, and the only ones that admit stable and fully homochiral states 
seem to be the ones which possess chiral inhibition reactions. We discuss in the next section some possible implications of that observation. 
\onecolumngrid

\begin{figure*}[h]
\centering
\includegraphics[scale=0.52]{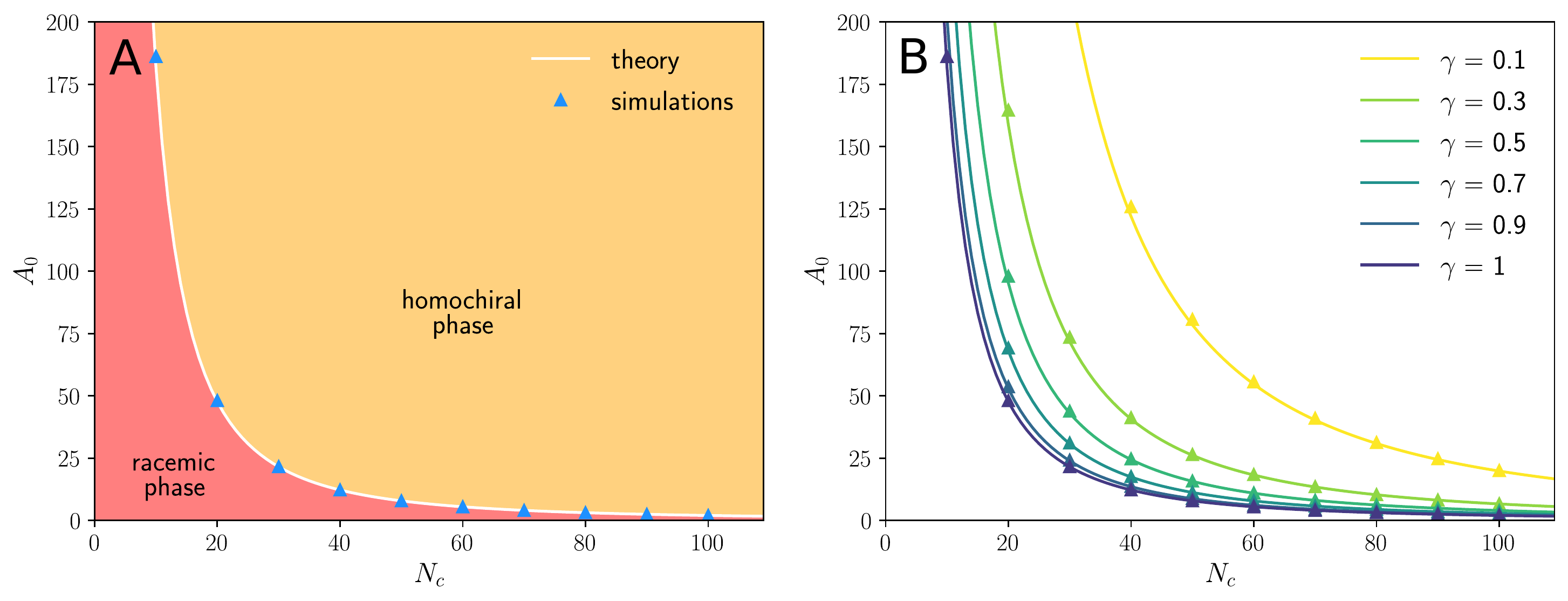}

\caption{Phase diagram showing the border between the racemic and homochiral phases in the $(A_0, N_{\rm C})$ plane for the generalized Frank model. $A_0$ is the inflow concentration of achiral species and $N_{\rm C}$ the number of chiral species in the reaction network. (A) In the lower domain of the diagram, the racemic state is stable against small perturbations. In the upper domain, the racemic state loses its stability while the homochiral state has gained stability. (B) Phase diagram showing the border between the racemic and homochiral phases for various sparsity coefficient $\gamma$. Below each solid line, the racemic phase is stable whereas above it it is unstable and instead the homochiral phase becomes stable. Simulations were performed for 100 realizations of the rate constants $k_{+}$, points represent simulations and solid lines the theory for both figures. The thresholds for both figures were computed by analyzing the stability of the racemic state of the system.}
\label{fig:phasediag}
\end{figure*}
\twocolumngrid

\subsection{Effect of sparsity on the reaction network}

Let us now explore the effect of sparsity on the stability of the reaction network. To model sparsity, we introduce a parameter $\gamma \in [0,1]$ which quantifies the probability of 
autocatalytic reactions [Eqs.~(\ref{react1}) and (\ref{react2})] to be actually present in the system. In other words, reactions of this type are assumed to have a zero rate constant with probability $1- \gamma$ and a non-zero value otherwise. Following the same reasoning as in the previous section, the threshold to reach the homochiral state then becomes $A_0 > A_0^*/\gamma$.
Thus, we find that the sparsity of the reaction network raises the threshold $A_0^*$, above which the homochiral phase is stable. This threshold $A_0^*$ is related to the amount of free energy brought into the system by incoming molecules. 
Note that we discuss here only the sparsity of autocatalytic reactions, which is distinct from the sparsity of the entire network. The latter was quantified by the parameter $\alpha$ in the random matrix argument. In any case, both forms of sparsity behave similarly, since both increase the thresholds to reach a homochiral state, where the thresholds may be either measured in terms of $A_0$ or in terms of the required number of chiral species $N_{\rm C}$, as illustrated in Fig. \ref{fig:phasediag}B.

Let us then introduce a second parameter, $\beta \in [0,1]$ that quantifies the probability of chiral inhibition reactions [Eq.~(\ref{react3})]. We find that, as long as every species (together with its enantiomer) is involved in at least one chiral inhibition reaction, the value of $\beta$ does not affect the stability of the homochiral state, while it affects the dynamics of the system. 
This can be understood from the expressions of the two largest eigenvalues $\lambda^L_{\rm max}$ and $\lambda_-$ deduced from 
Eqs.~(\ref{lambda_m}) and~(\ref{lambda_L-main}):
\be
\begin{multlined}
\lambda^L_{\rm max} = \beta \langle \tilde{k}_- \rangle  \left( \frac{1}{\gamma}\, A_0^* - A_0 \right) 
\qquad\mbox{and}\qquad \\
\lambda_- = \frac{1}{\tau} \left(1 - \frac{\gamma A_0}{A_0^*}\right) \, .
\end{multlined}
\ee
Note that the true maximal eigenvalue of ${\boldsymbol{\mathsf N}}$ is $\lambda_{\rm MAX} = \max( \lambda^L_{\rm max}, \lambda_-, \lambda_+)$. Thus the characteristic time for the system to converge towards the homochiral state when $A_0 > \frac{1}{\gamma} A_0^*$ is $t_c \sim 1/|\lambda_{\rm MAX}|$. Therefore, there are different regimes, and the characteristic time for the system to reach the homochiral state is
\be
t_c \sim \frac{1}{|\lambda^L_{\rm max}|} \propto \frac{1}{\beta \left(A_0^*/\gamma -A_0\right)}, 
\ee
if $\lambda^L_{\rm max} > \lambda_-$, which means that the dynamics becomes slower as the proportion of chiral inhibition reactions 
decreases in agreement with polymer models \cite{sandars_toy_2003},  and
\be
t_c \sim \frac{1}{|\lambda_-|} \propto \frac{\gamma}{A_0^*} \left(\frac{1}{\gamma}A_0^* - A_0 \right)
\ee
otherwise. Note that there is also a critical slowing down as $A_0$ approaches the threshold value $A_0^* / \gamma$, 
which was already present in non-sparse networks.


The emergence of homochirality present interesting analogies with the emergence of autocatalytic sets within a chemical network \cite{ribo_chemical_2019}, 
which has been a classic topic in the research on the origins of life since the pioneering work by Kauffman \cite{kauffman_autocatalytic_1986} recently reviewed in \cite{ameta_self-reproduction_2021,hordijk_history_2019}. 
In both transitions, the bifurcation scenario and the type of reaction network are similar and both transitions are robust with respect to background reactions because they depend primarily on the connectivity of the chemical network.

\section{Discussion}

In his monumental work on the origin of species \cite{darwin_1859}, Darwin imagined that life could have started in a 
warm little pond. While there is still no consensus about this, the idea of the warm little pond
presents many important aspects: 
first, it suggests an environment with  complex chemistry, second, it suggests a restricted physical space where a large number 
of chemicals could have been naturally concentrated as a result of changes in the environment  due for instance to evaporation or day-night cycles and, thirdly, it suggests a place where chemical species or structures may have to interact to access energy, either coming from sun light, from concentration gradients or from high-energy substrate molecules. 
In any case, such a setting seems more suitable than a fully open area like the middle of an ocean where dilution or draining by side reactions would have been major hurdles.  

Our scenario fully incorporates this idea of complex chemistry, because it is not based on any specific chemical network, and there is also a lot of freedom in the way it can be put out of equilibrium [thanks to a continuous stirred-tank reactor (CSTR) or to chemostats, or  to compartmentalization dynamics]. The main condition is that the system is sufficiently 'messy' so that its composition contains a large number of interacting species. In addition to the number of species, it is important to realize that two other control parameters can drive the transition as illustrated graphically in Fig. \ref{fig:phasediag}A and Fig. \ref{fig:phasediag}B. These two parameters are the strength of the interactions among species (which is controlled by the parameter $A_0$ in the generalized Frank model) and the sparsity of the network. Both parameters have also been identified as key parameters to control ecosystem stability \cite{ratzke_strength_2020}.

An important model system in this context is Viedma deracemization \cite{viedma_chiral_2005}, which is a process based on a breakage-fusion recycling of clusters combined with a racemization of the monomers in solution \cite{buhse_spontaneous_2021,blanco_mechanically_2017}. We note that many features of that process have analogs in our model: first, the racemization reaction of the monomers plays a similar role as the mutual inhibition reaction in Frank’s original model, although there are no true chiral cross-inhibition from hetero-chiral interactions. Secondly, the Viedma deracemization requires a polydispersity of cluster sizes, which is in line with the requirement of a large number of chiral species in our model. Thirdly, as mentioned before, the injection of energy is also required, by mechanical grinding in Viedma deracemization and by the injection of chemical species in the generalized Frank's model. The discovery of Viedma deracemization triggered promising applications for the enantioselective synthesis of certain compounds \cite{noorduin_emergence_2008}. On a more fundamental level, this work shows that homochirality can result from a competitive self-assembly of building-blocks into a large panel of chiral structures, which need not be regular and well formed polymers. If the conditions are right, even 'messy' aggregates can qualify.

Let us now discuss more general schemes which do not require the formation of soluble clusters or crystals. One attractive possibility is to link homochirality with the formation of the first polymers. 
In that case, theoretical studies have shown that the homochiral phase can be stabilized by chiral inhibition reactions \cite{sandars_toy_2003}.
Intuitively, this can be understood by the following simple argument. To investigate the stability of the homochiral state, it is sufficient to consider  a homochiral state of $N$ molecules to which one would add another chiral molecule. In the presence of inhibition reactions, this additional molecule would not be able to invade the already present homochiral cluster, because it would be blocked by these reactions.  As a result, the homochiral state would be necessarily stable.

To summarize, the emergence of homochirality appears to be driven by two forces: the multiplication of chiral species -- linked to the combinatorial explosion in the number of chiral species as molecules become  larger -- and chiral inhibition reactions. This leads to two remarks: the first one is that chiral inhibition would favor the emergence of chiral compartments of  different chirality by phase separation as observed in some experiments with coacervates. One might imagine that groups of molecules belonging to different compartments could then develop distinct functions (such as sugars and amino acids for instance). This further suggests that compartmentalization and homochirality could be somehow linked: in our previous works, we already noted that compartmentalization enhances the diversity of chemistry allowed by autocatalysis \cite{blokhuis_universal_2020} and effectively increases the number of chemical species \cite{laurent_emergence_2021}, which both favor a homochiral state. Experiments carried out with random RNA sequences also suggest a possible link between compartmentalization and homochirality because mixtures of such polymers can self-assemble into chiral liquid crystal domains when the polymers are sufficiently long \cite{bellini_liquid_2012}. 

The second point is that chiral inhibition naturally occurs in polymerization because it is well known that the addition of a monomer of the wrong handedness can bring polymerization to a halt \cite{joyce_chiral_1984}.  Therefore, it is possible that homochirality could have emerged together with the formation of the first biological polymers, possibly precursors to RNA \cite{tupper_role_2017}. Indeed, this is the time, where a multiplication of the number of species first occurred as a result of the formation of long chains due to combinatorial explosion \cite{bolli_pyranosyl-rna_1997}, and where cross-inhibition reactions became important. Theoretical polymerization models have confirmed that both features are required to reach full homochirality \cite{sandars_chirality_2005,gleiser_extended_2008}.


If this polymerization was taking place at equilibrium, the yield of long polymers would be very small and this would be an issue. Fortunately, these constraints can be overcome  
when the system is far from equilibrium, for instance when it is driven by temperature gradients \cite{baaske_extreme_2007}, or day-night cyles \cite{damer_coupled_2015}. In such a setting, long
polymers can be produced for instance thanks to template-assisted ligation, even in the presence of size-dependent severing and degradation reactions due to hydrolysis. 
Given these basic ingredients, the emergence of homochirality in the first polymers could have concerned all types of polymers: pre-RNA molecules and precursors of proteins or lipids, and this mechanism would also have selected specific relative signs between the different groups of molecules.

\subsection*{Acknowledgments}
The authors acknowledge fruitful discussions with L. Jullien, M. Yah Ben Zion and S. Walker. PG acknowledges the financial support of the Universit\'e Libre de Bruxelles (ULB) and the Fonds de la Recherche Scientifique-FNRS under the grant PDR T.0094.16 for the project ``SYMSTATPHYS''. DL acknowledges support from Agence Nationale de la Recherche (ANR-10-IDEX-0001-02, IRIS OCAV) and (ANR-11-LABX-0038, ANR-10-IDEX-0001-02).

\onecolumngrid
\appendix
\section*{Appendix A: Elements of the Jacobian matrix}
\label{AppA}

Here is the list of the elements of the Jacobian matrix ${\boldsymbol{\mathsf N}}$. The elements are evaluated in the fully homochiral state where $L_i = 0$, $\forall i$:
\begin{eqnarray}
	\frac{\partial \dot{A}}{\partial A} &=&  - \sum_{ijk \atop j\le k} k_{+ijk} \, D_i   - \frac{1}{\tau}\, , \qquad
	\frac{\partial \dot{A}}{\partial D_n} =  - \sum_{jk \atop j\le k} k_{+njk} \, A\, , \qquad
	\frac{\partial \dot{A}}{\partial L_n} =  - \sum_{jk \atop j\le k} k_{+njk} \, A\, , \qquad\quad \label{dapda} \\
	\frac{\partial \dot{D_m}}{\partial A} &=&  - \sum_{ij \atop i\le j} k_{+mij} \, D_m  + \sum_{ij \atop m\le j} k_{+imj} \, D_i + \sum_{ij \atop j\le m} k_{+ijm} \, D_i\, , \label{ddpda} \\
	\frac{\partial \dot{D_m}}{\partial D_n} &=&  \sum_{i \atop m\le i} k_{+nmi} \, A  + \sum_{i \atop i\le m}  k_{+nim}  \, A  -\delta_{nm}\Big(A\sum_{ij\atop i\le j} k_{+mij} +\frac{1}{\tau} \Big) \, , \label{eqn1}\\
	\frac{\partial \dot{D_m}}{\partial L_n} &=& -\tilde k_{-mn}D_m \, , \qquad\qquad
	\frac{\partial \dot{L_m}}{\partial A} =  0\, , \qquad\qquad
	\frac{\partial \dot{L_m}}{\partial D_n} = 0 \, ,\\
	\frac{\partial \dot{L_m}}{\partial L_n} &=&  \sum_{i \atop m\le i}  k_{+nmi}  \, A  + \sum_{i \atop i\le m}  k_{+nim}  \, A  -\delta_{nm}\Big(A\sum_{ij\atop i\le j} k_{+mij} + \sum_i\tilde k_{-im }D_i +\frac{1}{\tau} \Big) \, . \label{dlpdl}
\end{eqnarray}

Let us denote as $\lambda^D$ (resp. $\lambda^L$) the eigenvalues of the matrix $\frac{\partial \dot{{\boldsymbol{\mathsf D}}}}{\partial {\boldsymbol{\mathsf D}}}$ (resp.  $\frac{\partial \dot{{\boldsymbol{\mathsf L}}}}{\partial {\boldsymbol{\mathsf L}}}$), and as $\lambda_{\rm max}^D$ (resp. $\lambda_{\rm max}^L$) the corresponding maximal eigenvalue. 
Using methods introduced in section S5 of the Supp. Mat of \cite{laurent_emergence_2021}, we obtain from (\ref{eqn1}) and (\ref{dlpdl}) 
\be	
\lambda_{\rm max}^D = A\langle k_+ \rangle \frac{N_{\rm C}(N_{\rm C} +1)}{2}-\frac{1}{\tau} \, , \qquad
\lambda_{\rm max}^L = A\langle k_+ \rangle \frac{N_{\rm C}(N_{\rm C} +1)}{2}-\frac{1}{\tau} + S \, ,  \label{lambda_L}
\ee
where $S$ is defined as $S = -D N_{\rm C} \langle \tilde{k}_{-} \rangle$ and is due to the term $-\sum_i\tilde k_{-im }D_i$ in Eq.~(\ref{dlpdl}) for the $m^{th}$ diagonal element.
Using the mean-field expression for $D$ derived in Eq.~(\ref{MF_D}), and the expression for $A$ from 
Eq~(\ref{A0threshold}) we obtain
\be
\lambda_{\rm max}^D = 0 \, ,  \qquad 
\lambda_{\rm max}^L = S = - N_{\rm C} D \langle \tilde{k}_- \rangle
\label{lD=0}
\ee
with $\langle \tilde{k}_- \rangle$ the mean value of chiral inhibition reaction constants.

\section*{Appendix B: Spectrum of the upper-left block of the Jacobian matrix}
\label{AppB}

Let us define the upper-left block submatrix of the matrix as $\boldsymbol{\mathsf H}$,
\be
\boldsymbol{\mathsf H} = \begin{pmatrix}
	a & & {\mathbf v}^{\rm T} \\ {\mathbf u} & & \boldsymbol{\mathsf K} 	\end{pmatrix} \, ,
\ee
where
\be\label{definitions}
a = \frac{\partial \dot{A}}{\partial A} \, , \qquad
({\mathbf v}^{\rm T})_n =  \frac{\partial \dot{A}}{\partial D_n} \, ,\qquad
({\mathbf u})_m = \frac{\partial \dot{D_m}}{\partial A} \, ,\qquad
(\boldsymbol{\mathsf K})_{mn} = \frac{\partial \dot{D_m}}{\partial D_n} \, .
\ee
In the large $N_{\rm C}$ limit, the vector ${\mathbf u}$ converges towards a constant vector. Its elements are given by [see Eq.~(\ref{ddpda})]
\be
({\mathbf u})_m \underset{N_{\rm C} \to \infty}{=} DN_{\rm C} \frac{N_{\rm C} + 1}{2} \langle k_+ \rangle = (A_0 - A_0^*)\frac{N_{\rm C} + 1}{2} \langle k_+ \rangle \, , \label{vec_u} 
\ee
using Eq.~(\ref{MF_D}). All  the elements of ${\mathbf u}$ converge to the same value given by the previous equation. We can do the same evaluation for ${\mathbf v}^{\rm T}$ defined by Eq.~(\ref{definitions}) using Eq.~(\ref{A0threshold}):
\be ({\mathbf v}^{\rm T})_n \underset{N_{\rm C} \to \infty}{=} - A N_{\rm C} \frac{N_{\rm C} +1}{2}\langle k_+ \rangle = - \frac{1}{\tau} \, . \label{vec_vT}
\ee
Finally, we can compute the value of the scalar element $a$ defined by Eq.~(\ref{definitions}) in the $\NC \to \infty$ limit:
\be
a \underset{N_{\rm C} \to \infty}{=} -D \langle k_+ \rangle N_{\rm C}^2 \frac{N_{\rm C} + 1}{2} - \frac{1}{\tau} = -\frac{A_0}{A_0^* \tau} \, . 
\label{a-large-Nc}
\ee
after simplification with Eqs.~(\ref{A0threshold}) and (\ref{MF_D}).
We write the characteristic determinant of the matrix $\boldsymbol{\mathsf H}$ using the Schur complement of the block $a$, with $\lambda$ a complex variable such that $\lambda \notin \sigma_{\boldsymbol{\mathsf K}} = \{\lambda_1, \lambda_2, \dots , \lambda_{N_{\rm C}}\}$, a condition which guarantees the invertibility of $\boldsymbol{\mathsf K} - \lambda \boldsymbol{\mathsf I}$:

\begin{equation}
\det(\boldsymbol{\mathsf H} - \lambda \boldsymbol{\mathsf I}) =  [a - \lambda - {\mathbf v}^{\rm T}\cdot (\boldsymbol{\mathsf K} - \lambda \boldsymbol{\mathsf I})^{-1}\cdot{\mathbf u}] \, \det(\boldsymbol{\mathsf K} - \lambda \boldsymbol{\mathsf I}) \, ,
\label{schur_complement}
\end{equation}
with $\boldsymbol{\mathsf I}$ the identity matrix. At this point, it is important to recall  that, in the large $N_{\rm C}$ limit, the isolated eigenvalue of the matrix $\boldsymbol{\mathsf K} = \frac{\partial \dot{{\boldsymbol{\mathsf D}}}}{\partial {\boldsymbol{\mathsf D}}}$, which we here denote $\lambda_1$, is associated to the eigenvector $(1,\dots,1)^{\rm T}$. Therefore we see that ${\mathbf u}$ and ${\mathbf v}^{\rm T}$ are eigenvectors of matrix $\boldsymbol{\mathsf K}$ associated to $\lambda_1$, i.e., $\boldsymbol{\mathsf K} \cdot {\mathbf u} = \lambda_1 {\mathbf u}$, thus
\begin{equation}
(\boldsymbol{\mathsf K} - \lambda \boldsymbol{\mathsf I})^{-1} \cdot {\mathbf u} = \frac{1}{\lambda_1 - \lambda} \, {\mathbf u} \, .
\end{equation}
A similar development could be done by 
considering ${\mathbf v}^{\rm T}$ as a left-eigenvector of $\boldsymbol{\mathsf K}$.  Substituting this relation into Eq.~(\ref{schur_complement}), we get

\begin{equation}
\det(\boldsymbol{\mathsf H} - \lambda \boldsymbol{\mathsf I}) =\left( a - \lambda - \frac{{\mathbf v}^{\rm T}\cdot{\mathbf u}}{\lambda_1 - \lambda}\right) \det(\boldsymbol{\mathsf K} - \lambda \boldsymbol{\mathsf I}) 
= \frac{(a - \lambda)(\lambda_1 - \lambda) -{\mathbf v}^{\rm T}\cdot{\mathbf u}}{\lambda_1 - \lambda}\, \det(\boldsymbol{\mathsf K} - \lambda \boldsymbol{\mathsf I}) \, .
\end{equation}
We use the decomposition of $\det(\boldsymbol{\mathsf K} - \lambda \boldsymbol{\mathsf I})$ as a  polynomial function of roots $\sigma_{\boldsymbol{\mathsf K}} = \{ \lambda_1, \lambda_2, \dots, \lambda_{N_{\rm C}} \}$ (i.e., the eigenvalues of $\boldsymbol{\mathsf K}$)  to find
\begin{equation}
\det(\boldsymbol{\mathsf H} - \lambda \boldsymbol{\mathsf I}) = [(a - \lambda)(\lambda_1 - \lambda) -{\mathbf v}^{\rm T}\cdot{\mathbf u}]\, (\lambda_2 - \lambda) \times\dots \times (\lambda_{N_{\rm C}} - \lambda) \, .
\label{polynome_caracteristique}
\end{equation}
By continuity (since initially we had assumed that  $\lambda \notin \sigma_{\boldsymbol{\mathsf K}}$), we deduce from Eq.~(\ref{polynome_caracteristique}) that the spectrum of matrix $\boldsymbol{\mathsf H}$ is given by the $N_{\rm C} -1$ eigenvalues of $\boldsymbol{\mathsf K}$ distributed in the circle (from Girko's theorem), and two eigenvalues solutions of $(a - \lambda)(\lambda_1 - \lambda) -{\mathbf v}^{\rm T}\cdot{\mathbf u} = 0$. As a result, the isolated eigenvalue $\lambda_1$ is modified and becomes one of the two solutions of the previous equation to solve. From Eq.~(\ref{lD=0}), the isolated eigenvalue $\lambda_1$ should be identified with the largest eigenvalue $\lambda_{\rm max}^D$. This means that $\lambda_1 \to 0$ in the large $N_{\rm C}$ limit. The equation that remains to be solved to fully determine the spectrum of $\boldsymbol{\mathsf H}$ is then
\begin{equation}
\lambda^2 - a \lambda - {\mathbf v}^{\rm T}\cdot{\mathbf u}=0 \, .
\label{eigenvalues_equation}
\end{equation}
Using the expressions  (\ref{vec_u}) and (\ref{vec_vT}) for ${\mathbf u}$ and ${\mathbf v}^{\rm T}$, as well as Eq.~(\ref{A0threshold}), we determine their dot product
\begin{equation}
{\mathbf v}^{\rm T}\cdot {\mathbf u} = - \frac{A_0 - A_0^*}{\tau}  N_{\rm C} \frac{N_{\rm C} + 1}{2} \langle k_+ \rangle =   -  \frac{A_0 - A_0^*}{\tau^2 A_0^*} \, ,
\end{equation}
where  the additional $N_{\rm C}$ factor comes from the summation.
With Eq.~(\ref{a-large-Nc}), the discriminant of Eq.~(\ref{eigenvalues_equation}) can thus be written
\be
\Delta = a^2 + 4\,{\mathbf v}^{\rm T}\cdot{\mathbf u} = \frac{1}{\tau ^2} \left(\frac{A_0}{A_0^*} - 2 \right)^2 \, .
\ee
We  observe that $\Delta > 0$ when $A_0 > A_0^*$,  but $\Delta = 0$ in the particular case where $A_0 = 2A_0^*$. When $A_0 \neq 2A_0^*$, the two real solutions of Eq.~(\ref{eigenvalues_equation}) are

\be
\lambda_+ = \frac{1}{2} \left[ -\frac{A_0}{A_0^* \tau} + \frac{1}{\tau}\left(\frac{A_0}{A_0^*} -2\right) \right] = - \frac{1}{\tau} \, ,
\ee
and
\be
\lambda_- = \frac{1}{2} \left[ -\frac{A_0}{A_0^* \tau} - \frac{1}{\tau}\left(\frac{A_0}{A_0^*} -2\right) \right] = \frac{1}{\tau} \left(1-\frac{A_0}{A_0^*} \right) \, .
\ee
Finally, in the particular case where $A_0 = 2A_0^*$, there is a unique solution, $\lambda_+ = \lambda_- = -1/\tau$.

\section*{Appendix C: Reversible and irreversible generalized Frank models}
\label{AppC}

	The kinetic equations of the reversible generalized Frank model (\ref{react1})-(\ref{react3}) with ${\rm E}_m={\rm D}_m$ and $\bar{\rm E}_m={\rm L}_m$ are given in an open CSTR by Eq.~(\ref{kin_eqs}) for the concentrations ${\bf c}=(\{A_a\}_{a=1}^{\NA},\{D_i,L_i\}_{i=1}^{\NC},\{\tilde A_a\}_{a=1}^{\NAi})$.  The explicit form of these equations is presented in the Supplementary Information of Ref.~\cite{laurent_emergence_2021}.
	
	In a closed system with an infinite mean residence time $\tau=\infty$, the concentrations are converging towards their unique stationary equilibrium values ${\bf c}_{\rm eq}$, at which the detailed balance conditions hold.  These latter are implying that the reaction rate constants are satisfying the following relations for $j \leq k$ and $b \leq c$ :
\be
k_{-aijk} = k_{+aijk} \, \frac{D_i\, A_a}{D_j\, D_k}\Bigg\vert_{\rm eq} = k_{+aijk} \, \frac{L_i\, A_a}{L_j\, L_k}\Bigg\vert_{\rm eq} \, , \qquad
\tilde k_{+bcij} = \tilde k_{-bcij} \, \frac{D_i\, L_j}{\tilde A_b \, \tilde A_c}\Bigg\vert_{\rm eq} \, ,
\ee
	where $\tilde k_{\pm bcij}=\tilde k_{\pm bcji}$ because of chiral symmetry, $k_{+aijk}$ ({\it resp.} $k_{-aijk}$) are the forward ({\it resp.} backward) rate constants of reactions (\ref{react1})-(\ref{react2}) and $k_{-bcij}$ ({\it resp.} $k_{+bcij}$) are the forward ({\it resp.} backward) rate constants of reactions (\ref{react3}). The direct consequence is that the rate constants are related to each other by compatibility conditions for the existence of equilibrium.  In the particular case where $\NC=2$ and $\NA=\NAi=1$, the compatibility conditions read
	\bea
	&& \frac{D_1}{A}\Bigg\vert_{\rm eq} = \frac{L_1}{A}\Bigg\vert_{\rm eq} = \frac{k_{+111}}{k_{-111}} = \frac{k_{+212}}{k_{-212}}= \sqrt{ \frac{k_{+211}}{k_{-211}}\, \frac{k_{+222}}{k_{-222}} } \, , \\
	&& \frac{D_2}{A}\Bigg\vert_{\rm eq} = \frac{L_2}{A}\Bigg\vert_{\rm eq} = \frac{k_{+222}}{k_{-222}} = \frac{k_{+112}}{k_{-112}}= \sqrt{ \frac{k_{+111}}{k_{-111}}\, \frac{k_{+122}}{k_{-122}} } \, , \\
	&& \frac{\tilde A}{A}\Bigg\vert_{\rm eq} =  \frac{k_{+111}}{k_{-111}} \sqrt{\frac{\tilde k_{-11}}{\tilde k_{+11}}} = \frac{k_{+222}}{k_{-222}} \sqrt{\frac{\tilde k_{-22}}{\tilde k_{+22}}}= \sqrt{\frac{k_{+111}}{k_{-111}}\, \frac{k_{+222}}{k_{-222}} \, \frac{\tilde k_{-12}}{\tilde k_{+12}} } \, , 
	\eea
	where $\tilde k_{\pm 12}=\tilde k_{\pm 21}$.  A further consequence is that the equilibrium state is always racemic.
	Another observation is that the equilibrium concentrations of the different species can be calculated in terms of the rate constants combined with the conservation law $(d/dt)[\sum_{a=1}^{\NA} A_a + \sum_{i=1}^{\NC}(D_i+L_i)+\sum_{a=1}^{\NAi} \tilde A_a]=0$, holding in the closed system.  
	
	We note that, if the rate constants have equal values according to $k_{\pm aijk}=k_{\pm}$ and $\tilde k_{\pm bcij}=\tilde k_{\pm}$ (which corresponds to the mean-field approximation) and if the initial concentrations in the closed system are taken as $A_a=A_0$, $D_i=0$, $L_i=0$, and $\tilde A_a=0$, the equilibrium concentrations are given by
	\be
	D_{\rm eq}=L_{\rm eq} = A_{\rm eq} \, \frac{k_+}{k_-} \, , \qquad 
	\tilde A_{\rm eq} = A_{\rm eq}  \, \frac{k_+}{k_-}  \, \sqrt{\frac{\tilde k_-}{\tilde k_+}} \, , \qquad 
	\mbox{with} \qquad
	A_{\rm eq} = \frac{A_0}{1 + 2\frac{\NC}{\NA} \frac{k_+}{k_-} + \frac{\NAi}{\NA}\frac{k_+}{k_-} \sqrt{\frac{\tilde k_-}{\tilde k_+}} } \, .
	\ee
	Consequently, in the limit where $k_- \ll k_+$ and $\tilde k_+ \ll \tilde k_-$, we find
	\be
	A_{\rm eq} \simeq A_0 \, \frac{\NA}{\NAi}\frac{k_-}{k_+} \sqrt{\frac{\tilde k_+}{\tilde k_-}} \, ,
	\qquad
	D_{\rm eq}=L_{\rm eq} \simeq A_0 \,  \frac{\NA}{\NAi}\sqrt{\frac{\tilde k_+}{\tilde k_-}} \, ,
	\qquad
	\tilde A_{\rm eq} \simeq  A_0  \,  \frac{\NA}{\NAi} \, ,
	\ee
	so that $\tilde A_{\rm eq} \gg D_{\rm eq}=L_{\rm eq} \gg A_{\rm eq}$ in agreement with the free energies assumed for the achiral species.  
	
	These considerations are showing that the rate constants $k_{-aijk}$ and $\tilde k_{+bcij}$ can take arbitrarily small values with respect to $k_{+aijk}$ and $\tilde k_{-bcij}$, which has for consequence that $\{A_{a,{\rm eq}}\}_{a=1}^{\NA} \ll \{\tilde A_{a,{\rm eq}}\}_{a=1}^{\NAi}$ in relation to the higher free energy for the achiral species $\{{\rm A}_a\}_{a=1}^{\NA}$ than for the product species $\{\tilde {\rm A}_a\}_{a=1}^{\NAi}$.  Therefore, the rate constants $k_{+aijk}$ and $\tilde k_{-bcij}$ may be supposed to be independent parameters.
	
	\vskip 0.5 cm
	
	In far-from-equilibrium open systems with a finite value of the mean residence time $0<\tau<\infty$, the forward reactions,
	\bea
	&&{\rm A}_a +{\rm D}_i  \ {\overset{k_{+aijk}}{\longrightarrow}} \ {\rm D}_j + {\rm D}_k \qquad (j\le k)\, , \label{react1bis} \\
	&&{\rm A}_a +{\rm L}_i  \ {\overset{k_{+aijk}}{\longrightarrow}} \  {\rm L}_j + {\rm L}_k \qquad (j\le k)\, , \label{react2bis} \\
	&&{\rm D}_i + {\rm L}_j   \ {\overset{\tilde k_{-bcij}}{\longrightarrow}} \  \tilde{\rm A}_b + \tilde{\rm A}_c  \qquad (b\le c)\, , \label{react3bis} 
	\eea
	(where $a=1,2,\dots,\NA$; $b,c=1,2,\dots,\NAi$; and $i,j,k=1,2,\dots,\NC$) are thus playing a dominant role and the reversed reactions can be negligible.  Accordingly, in these far-from-equilibrium regimes, the fully irreversible generalized Frank model may be considered, as discussed in Ref. \cite{laurent_emergence_2021}.  The kinetic equations of the fully irreversible model (\ref{react1bis})-(\ref{react3bis}) are explicitly given by equation set (\ref{eq-A-num2}).  For the fully irreversible model, the rate constants $k_{+aijk}$ and $\tilde k_{-bcij}$ can thus take independent values fixed according to the experimental observations.

\twocolumngrid
\bibliographystyle{ieeetr}

\end{document}